\documentclass[a4paper,11pt]{article}
\pdfoutput=1
\usepackage{jheppub}
\usepackage{xcolor}

\usepackage{latexsym,amsbsy}
\usepackage{slashed}

\usepackage[compat=1.1.0]{tikz-feynman}

\newcommand{\BE}{\begin{equation}}
\newcommand{\EE}{\end{equation}}
\newcommand{\BQ}{\begin{equation} \begin{array}{c}}
\newcommand{\EQ}{\end{array}\end{equation}}
\newcommand{\BQA}{\begin{equation} \begin{aligned}}
\newcommand{\EQA}{\end{aligned}\end{equation}}
\newcommand{\BT}{\begin{theorem}}
\newcommand{\ET}{\end{theorem}}

\newcommand{\bc}{\begin{center}}
\newcommand{\ec}{\end{center}}

\newcommand{\FHS}{\breve{F}}

\newcommand{\PhiB}{\overline{\Phi}}
\newcommand{\BB}{\overline{B}}
\newcommand{\dT}{\widetilde{d}}
\newcommand{\AT}{\widetilde{A}}

\newcommand{\ATS}{\widetilde{\slashed{A}}}
\newcommand{\DTS}{\widetilde{\slashed{D}}}

\newcommand{\DT}{\widetilde{D}}
\newcommand{\FT}{\widetilde{F}}

\newcommand{\lX}{\lambda}

\newcommand{\LAG}{\mathcal{L}}
\newcommand{\ZZ}{\mathbb{Z}}
\newcommand{\QQ}{\mathbb{Q}}
\newcommand{\CC}{\mathbb{C}}

\newcommand{\demi}{{\frac{1}{2}}}

\newcommand{\SG}{{\sigma}}
\newcommand{\SB}{{\overline{\sigma}}}
\newcommand{\psiB}{{\overline{\psi}}}

\pgfkeys{/tikzfeynman/warn luatex=false}
\makeatletter\def\tikzfeynman@luatex@required@path{\relax}\makeatother

\begin{document}

\title {
  SU(2/1) superchiral self-duality: a new quantum, algebraic and geometric paradigm to describe the electroweak interactions.
  }


\author{Jean Thierry-Mieg${}^1$, Peter Jarvis${}^{2,3}$.}

\affiliation{${}^{1}$NCBI, National Library of Medicine, National Institute of Health, \\
  8600 Rockville Pike, Bethesda MD20894, U.S.A. \\
  ${}^{2}$School of Natural Sciences (Mathematics and Physics),\\
  University of Tasmania, Private Bag 37,\\
  Hobart, Tasmania 7001, Australia\\
  ${}^3$Alexander von Humboldt Fellow.}

\emailAdd{mieg@ncbi.nlm.nih.gov, peter.jarvis@utas.edu.au}

\abstract{
We propose an extension of the Yang-Mills paradigm from Lie algebras to internal chiral superalgebras. We replace the Lie algebra-valued connection one-form $A$, by a superalgebra-valued polyform $\widetilde{A}$ mixing exterior-forms of all degrees and satisfying the chiral self-duality condition $\widetilde{A} = {}^*\widetilde{A} \,\chi$, where $\chi$ denotes the superalgebra grading operator. This superconnection contains Yang-Mills vectors valued in the even Lie subalgebra, together with scalars and self-dual tensors valued in the odd module, all coupling only to the charge parity CP-positive Fermions. The Fermion quantum loops then induce the usual Yang-Mills-scalar Lagrangian, the self-dual Avdeev-Chizhov propagator of the tensors, plus a new vector-scalar-tensor vertex and several quartic terms which match the geometric definition of the supercurvature. Applied to the $SU(2/1)$ Lie-Kac simple superalgebra, which naturally classifies all the elementary particles, the resulting quantum field theory is anomaly-free and the interactions are governed by the super-Killing metric and by the structure constants of the superalgebra.
}

\maketitle


\flushbottom

\section {Introduction}
The weak interactions are chiral. Before symmetry breaking, all the Fermions
of the standard model are massless, all the left states are $SU(2)$ doublets and all
the right states are singlets.
This fundamental asymmetry is difficult to justify in the Yang-Mills framework 
because Lie algebra symmetries can only connect states of a given chirality,
and connecting left particles to left antiparticles as in the $SU(5)$ grand-unified theory potentially implies
proton decay.
However, as observed in 1979 by Ne'eman  \cite{N1} and Fairlie \cite{F1}, the $SU(2)U(1)$ electroweak
algebra is naturally embedded in $SU(2/1)$, the smallest Lie-Kac
simple superalgebra \cite{Kac1}. The leptons \cite{N1,F1} and quarks \cite{DJ,NTM1},
graded by their chirality, fit the smallest irreducible representations
of $SU(2/1)$. Strangely, these representations are non Hermitian (\cite{TM20b}, appendix D). But we noticed
recently \cite{TM20b} that the resulting collection of scalar anomalies cancels whenever the
Adler-Bell-Jackiw \cite{Adler,BJ} vector anomaly cancels \cite{BIM}. These observations renew the interest in the
construction of a generalization of the Lie algebra Yang-Mills framework to the case
of a chiral superalgebra.

Merging differential geometry, superalgebra and quantum field theory
concepts, we present a new paradigm. We propose to
consider as fundamental
a Lie superalgebra-valued superconnection polyform $\AT$, 
mixing de Rham exterior-forms of all degrees \cite{NTM82,Quillen,MQ86,TM20a} 
and satisfying the new superchirality condition:
\BE \label{eq1}
\AT = ^*\AT\;\;\chi\;,
\EE
where the $^*$ denotes the Hodge duality in Minkowski 4-dimensional space-time
with signature $(-+++)$ and $\chi$ is the charge-chirality of the superalgebra instrumental in the
definition of the supertrace
\BE \label{eq2}
   STr (M) = Tr (\chi\;M)\;. 
\EE

Remarkably, this superchirality condition pairs charge conjugation
with parity, naturally enforcing the Landau charge parity (CP) invariance characteristic of the weak interactions.
The pairing is a consequence of the duality identities structuring the
antisymmetrized products of the Pauli matrices, which imply that if
$\AT$ is a self-dual (or anti-self-dual) polyform, then the corresponding Dirac-Yukawa operator
$\ATS$ only couples to left spinors $\psi_L$ (or right spinors $\psi_R$):
\BE
\AT = ^*\AT \Rightarrow  \ATS\;\psi_R = 0\;,\;\;\,  \AT = -^*\AT \Rightarrow \ATS\;\psi_L = 0\;.
\EE
Expanding (1.1) in terms of the underlying fields,
we find (2.8) that the vectors couple exactly as postulated in 1979 by Ne'eman \cite{N1} and Fairlie \cite{F1},
that the scalars $\PhiB\Phi$ (2.9) couple exactly as in Thierry-Mieg \cite{TM20b}
and that 2-form components $\BB B$ of the superconnection, interpreted
as self and anti-self-dual Avdeev-Chizhov fields \cite{AC94},
follow the same pattern (2.10).
A new trilinear scalar-vector-tensor interaction $F \{\BB,\Phi\}$
is induced by the Fermion loop  (4.8). It must be considered as an intrinsic part
of the minimal coupling of a superalgebra since the same term appears
in the square of the supercurvature defined in (2.11).

Curiously, the Fermion quantum loop counterterms hesitate between a Lie algebra
and a Lie superalgebra structure (see equations (3.7), (4.6), (5.1) and (5.2))\,, but when the construction is applied
to the $SU(2/1)$ model of leptons and quarks \cite{TM20b},
a generalization of the Bouchiat, Iliopoulos and Meyer (BIM) mechanism \cite{BIM}
lifts the ambiguity (section 7) and implies that
the theory is anomaly-free and that the propagators and
covariant derivatives of the scalars and the tensors are provided by the
antisymmetric $g_{ij}$ part of the super-Killing metric of
the superalgebra, and by the symmetric part $d^{a}_{ij}$ of its structure constants.

From a geometrical perspective, our new complementary treatment of the
exterior bundle (the polyform superconnection) with the spinor bundle
(the chiral Fermions) reflects an element of \textit{bona fide} internal supersymmetry.
The signs generated in the quantum loops by the tensorial structures of
the propagators and by the orientations of the chiral Fermion propagators
match the signs generated by the grading of the superalgebra in the
Clebsch-Gordan calculations and their interplay with the super-Jacobi identity.
The theory is superalgebraic despite the fact that all the gauge fields are Bosons
and all the matter fields are Fermions, as requested by the
spin-statistics relation.

The theory respects Einstein's distinction
between the force fields, which are geometrized as superalgebra-valued polyforms,
and matter fields, represented by pointlike chiral Fermions. The odd couplings
play with the orientation of space, which is represented by the opposite
helicities of the massless left/right Fermions, and with the orientation of the p-forms matter fields and their Hodge duals.
This is very different from
the Wess-Zumino supersymmetry which couples Bosons to Fermions, i.e. force fields to matter fields.
Although the latter is much more developed, one should remember that not a single known particle is
the supersymmetric partner of another known particle -- for example the neutrino is not the partner
of the photon, whereas on the contrary, all the known elementary particles, the leptons and the quarks, naturally fall
in superchiral $SU(2/1)$ multiplets. Finally, the superconnection offers a way to geometrize
the Higgs fields.

In section 2, we introduce in details the new paradigm of a self superchiral superconnection.
In sections 3 and 4, we show how the knowledge of the couplings of the Fermions to the
scalar, vector and 2-tensor components of the superconnection
induce their propagators and interactions.
In section 5 to 7, we analyze the quantum anomalies.
Notations are given explicitly in appendix A to E.

\section{The new superchiral superconnection $\AT$}

The de Rham complex over a 4-dimensional differentiable manifold is the
space of all differential exterior-forms of all degrees from 0 to 4: $\AT = \phi + a + b + c + e$.
In Yang-Mills theory, the scalars $\phi$ are considered as zero-forms, i.e.
ordinary functions, and the Yang-Mills vector $a_{\mu}$ can be identified with the
components of a Lie algebra-valued Cartan connection one-form $a = a^a_{\mu}\lX_adx^{\mu}$
where the $\lX_a$ are the generators of the Lie algebra. The connection $a$ 
defines the parallel transport on the manifold and specifies how to rotate the fields
in internal space under an infinitesimal displacement in the base space by replacing the Cartan exterior
differential $d$ by the covariant exterior differential $D = d + a$.
Since exterior-forms of even degree ($\phi,b,e)$ commute, and exterior-forms of odd degree $(a,c)$
anticommute, it is natural, in extending the formalism to the superalgebra case (see Ne'eman-Thierry-Mieg \cite {NTM82}, Quillen \cite {Quillen,MQ86}) to associate the $\ZZ_2$ grading of the exterior-forms
to the $\ZZ_2$ grading of the superalgebra (appendix A), and to try to define a superconnection as
a globally odd form, that is, to keep only the odd exterior-forms of degree 1 and 3 
which are valued only in the even Lie subalgebra, 
$a + c =(a^a_{\mu}dx^{\mu} + c^a_{\mu\nu\rho}dx^{\mu}dx^{\nu}dx^{\rho}/6)\lX_a$\,,
together with the even forms of degree 0, 2 and 4,
$\phi +b+e=\phi^i + b^i_{\mu\nu}dx^{\mu}dx^{\nu}/2 +e^i_{\mu\nu\rho\sigma}dx^{\mu}dx^{\nu}dx^{\rho}dx^{\sigma}/24)\lX_i$\,,
which are valued in the odd module of the superalgebra.A. In \cite{TM20a}, we have shown that this definition
is incomplete, because the odd and even forms commute, $\phi^ia^a = a^a\phi^i$, whereas at the component level, we need (A.3) to
generate the antisymmetric commutator of the even and odd matrices. The paradox is resolved by
invoking the superalgebra charge chirality matrix $\chi$ (see the details in appendix A), which defines the supertrace of the superalgebra, commutes with
the even matrices, and anticommutes with the odd matrices:
\BE \label{eq3}
STr (M) = Tr (\chi\;M)\;,\;\;[\chi,\lX_a] = 0\;,\;\;\{\chi,\lX_i\} = 0\;.
\EE
Our final definition of the superconnection is
\BQ
\AT = (\phi + b + e)^i\lX_i + \chi \;(a + c)^a\lX_a\;,
\\
\dT = \chi\,d\;,\;\;\DT = \dT + \AT\;,\;\;\FT = \dT\AT + \AT\AT\;.
\EQ
The presence of the superalgebra-grading-matrix $\chi$ ensures that the signs arising in the construction
of the curvature polyform $\FT$, and in the action of $\DT$ on all fields, are always consistent
with the brackets and structure relations of the superalgebra \cite{TM20a}. As a result, the curvature $\FT$ defined as the square
of the covariant differential $\FT = \DT\DT$ is valued
in the adjoint representation of the superalgebra, defines a linear map,
and satisfies the Bianchi identity $\DT\FT = 0$, which in turn implies that the covariant differential is associative $(\DT\DT)\DT = \DT(\DT\DT)$. This geometric construction is satisfactory, but it does not yet explain the structure of the electroweak interactions.

The new concept presented here is, firstly, to consider in Minkowski 4-dimensional space-time with signature $(-+++)$,
a self-dual superconnection $\AT = ^*\AT$, where $^*$ denotes the Hodge duality which maps $p$-forms onto $(4-p)$-forms (appendix C).
In Yang-Mills theory, the connection $a$ is a 1-form, its dual $^*a$ is a 3-form, so a Yang-Mills connection cannot be self-dual.
We are only familiar with
the self-dual topological theories satisfying $F = ^*F$. But because a superconnection is composed of
exterior-forms of all degrees, its 1-form component $a$ can be the dual $a = ^*c$ of its 3-form component $c$
and the concept of a self-dual superconnection makes sense.

This constraint has a remarkable consequence when we consider the action of the superconnection
on chiral spinors. To construct this action, we saturate the Lorentz indices of the component
$p$-forms with Dirac $\gamma$ matrices, effectively defining a map in spinor space
using the Dirac-Feynman slash operator. The classic Dirac mapping
$a = a_{\mu}dx^{\mu}\;\Rightarrow\slashed{a} = a_{\mu} (\SG^{\mu} + \SB^{\mu})$ is generalized to antisymmetric tensors of
any rank, for example $b = \demi b_{\mu\nu}dx^{\mu}dx^{\nu}\;\Rightarrow\slashed{b} = \demi b_{\mu\nu}(\SB^{\mu}\SG^{\nu}+\SG^{\mu}\SB^{\nu})$.
As all our spinors are chiral, we use the $\gamma_5$ diagonal notation $\gamma_{\mu} (1+\gamma_5)/2 + \gamma_{\mu} ((1 - \gamma_5)/2 \rightarrow \SG_{\mu} + \SB_{\mu}$ as explained in appendix B.
However, the anti-symmetric product of $p$ Pauli
matrices can be rewritten as a product of $4 - p$ Pauli matrices contracted with
the antisymmetric Levi-Civita $\epsilon$ symbol (B.7). Therefore, the Dirac operator associated to a $p$-form $\omega$ can be rewritten as $\pm$ the Dirac operator
associated to its Hodge dual $^*\omega$, where the sign depends on the helicity of the two component Fermions on which we act (C.8).
For example, if a 3-form $c$ acts on left Fermions, this can be expressed in terms of the dual 1-form $^*c$ (C.6) :
\BE
\slashed{c} \; \frac {1 - \gamma_5}{2} = \frac{1}{6} c_{\mu\nu\rho}\SB^{\mu}\SG^{\nu}\SB^{\rho} = \frac{i}{6} c_{\mu\nu\rho}\epsilon^{\mu\nu\rho\sigma} \SB_{\sigma} = (^*c)_{\mu}\SB^{\mu}
\EE
Applying this transformation to the 2, 3 and 4 forms ($b,c$ and $e$)\,, the Dirac operator associated to the superconnection $\AT$ acting on the left Fermions can be rewritten as
\BE
\ATS \; \frac {1 - \gamma_5}{2} = (\phi + ^*e) \frac {1 - \gamma_5}{2}+ (a + ^*c)_{\mu}\;\SB^{\mu} + \frac{1}{2}\;(b + ^*b)_{\mu\nu}\;\SG^{\mu}\SB^{\nu}  
\;,
\EE
whereas the Dirac operator associated to the superconnection acting on the right Fermions can be rewritten as
\BE
\ATS \; \frac {1 + \gamma_5}{2} = (\phi - ^*e) \frac {1 + \gamma_5}{2}+ (a - ^*c)_{\mu}\;\SG^{\mu} + \frac{1}{2}\;(b - ^*b)_{\mu\nu}\;\SB^{\mu}\SG^{\nu}  \;.
\EE
Each parenthesized term pairs a $p$-form to the dual of the matching $(4-p)$-form.
As a result, see the details in appendix C, a self-dual superconnection annihilates the right Fermions and \textit{mutatis mutandis} an anti-self-dual
superconnection annihilates the left Fermions
\BE
\AT = ^*\AT \Rightarrow \ATS\;\psi_R = 0\;,\;\;\,  \AT = -^*\AT \Rightarrow \ATS\;\psi_L = 0\;.
\EE

To describe the electroweak interactions, we need to act both on left and on right Fermions, but with
different kinds of forces. In a superalgebra framework, the charge chirality operator $\chi$
 (2.1) that we have already introduced in the definition (2.2) of the superconnection provides this
distinction and we postulate that our superconnection should, in addition, be superchiral
\BE
\AT = ^*\AT\;\chi\;.
\EE
This beautiful equation (1.1)
correlates the orientation of space, which is hidden in the definition of
the Hodge duality, denoted by the $^*$, to the charge  
chirality $\chi$ of the superalgebra, defined in the internal charge space.
In consequence it constrains the chirality of the charged Fermions.

We illustrate the outcome of these constraints on the specific case of $SU(m/n)$ viewed as a chiral superalgebra.
In the $SU(m)$ sector, the potential (in the $SU(m/n)$ fundamental representation)
is accompanied by the sign $\chi = +1$, so $a = ^*c$ (C.7). Hence, in the Dirac operator,
only the term $(a + ^*c)_{\mu}\;\SB^{\mu}$ survives (see (C.6)-(C.8)) and it acts
only on the left Fermion, (B.3), (C.9) and (2.6). Reciprocally, in the $SU(n)$
sector ($\chi=-1$) , $\AT$ is anti-self-dual and the Dirac operator $(a - ^*c)_{\mu}\;\SG^{\mu}$
annihilates the left Fermions and only only on the right Fermions.

The $U(1)$ operator of $SU(m/n)$ is special in that the corresponding
matrix in the fundamental representation acts at the same time on $\chi = 1$ and $\chi = -1$
states, and satisfies $a = ^*c \chi$. In consequence, the supertraceless $U(1)$ Abelian vector
acts both on the left and right Fermions
via $a_{\mu}  ((1 + \chi)\SB^{\mu} + (1 - \chi)\SG^{\mu})$.

Returning to the case of $SU(2/1)$, we get $diagonal(\SB,\SB;2\,\SG)$,
exactly as postulated \textit{ex nihilo} in 1979 by Ne'eman \cite{N1} and Fairlie \cite{F1}
\BQ
  \slashed{A} = \frac{1}{4}\;A^a_{\mu}\lX_a\;(\SB^{\mu}(1+\chi)(1-\gamma_5) + \SG^{\mu} (1-\chi)(1+\gamma_5))\;.
\EQ
For the scalar fields, we have $\PhiB = \phi + ^*e$ and $\Phi = \phi - ^*e$
which act as
\BQ
\PhiB = \frac{1}{4}\;\PhiB^i\lX_i\;(1+\chi)(1-\gamma_5) 
\;,\\
\Phi = \frac{1}{4}\;\Phi^i\lX_i\;(1-\chi)(1+\gamma_5)
\;,
\EQ
exactly as we postulated \textit{ex nihilo} in \cite{TM20b}. The 2-form $b$ follows a similar pattern.
Separating the self-dual and anti-self-dual parts $\BB = b + ^*b$ and $B = b - ^*b$,
the Dirac operator acts as
\BQ
\slashed{\BB} = \frac{1}{8}\;\BB^i_{\mu\nu}\lX_i\;\SG^{\mu}\SB^{\nu}\;(1+\chi)(1-\gamma_5) 
\;,\\
\slashed{B} = \frac{1}{8}\;B^i_{\mu\nu}\lX_i\;\SB^{\mu}\SG^{\nu}\;(1-\chi)(1+\gamma_5)
\;.
\EQ
$\Phi$ and $B$ absorb right Fermions and emit left Fermions, and their antiparticles
$\PhiB$ and $\BB$ absorb left Fermions and emit right Fermions, as illustrated
below in the Feynman diagrams of sections 3 and 4.
Our point is that the superchiral constraint allows us to derive from first principles
the interactions that had to be imposed \emph{ad hoc} in the previous $SU(2/1)$ literature,
to force the gauge superalgebra to look like the standard model.
The price we pay is the appearance of a new scalar sector
represented by the $\BB B$ fields.

The reader should notice that the 2-form component $\FHS$ of the curvature polyform $\FT$ (2.2) reads, in these
new notations,
\BE
\FHS = (dA^a + \frac{1}{2} (f^a_{bc}A^bA^c + d^a_{ij} (\PhiB^i B^j + \Phi^i \BB^j)))\;\lX_a\;.
\EE
This generates inside the Lagrangian $\FT^2$ a new scalar-vector-tensor interaction $F (\{\BB,\Phi\} + \{\PhiB, B\})$.
As shown below, this term plays a crucial role in the self-consistency of the theory.

Given these algebraic and geometric definitions, let us now study how the
Dirac action of the superconnection on the chiral Fermions
gets promoted in the quantum field theory into the definition of the
propagators and interactions of its components, the complex scalar field $\PhiB\Phi$, the vector $A$,
and the complex self-dual anti-self-dual antisymmetric tensor $\BB B$, 
all correctly satisfying the spin-statistics relation.

\section {The Avdeev-Chizhov propagator is induced by the Fermion loop}

In their seminal study \cite{AC94}, Avdeev and Chizhov have introduced a new type of quantum fields:
a self-dual and an anti-self-dual antisymmetric tensor $\BB$  and $B$
satisfying in Minkowski space the conditions
\BE
\BB =  ^*\BB \;,\;\;\;B =  - ^*B\;,
\EE
where $^*$ denotes Hodge duality (C.1) in 4-dimensional Minkowski space-time with signature $(-+++)$:
\BQ
B = \frac {1}{2} B_{\mu\nu}\;dx^{\mu}dx^{\nu} \;,\;\;^*B = -\frac{i}{2}\;\epsilon_{\mu\nu\rho\sigma} B^{\mu\nu}\;dx^{\rho}dx^{\sigma}
\EQ
and $\epsilon$ is the fully antisymmetric Levi-Civita symbol with $\epsilon_{0123} = 1$.
These fields coincide with the antisymmetric tensor fields identified in (2.10) as part of the
superchiral superconnection $\AT$. Compare (3.2) with (C.1,C.5) and the definition of the Hodge dual of
the field components (C.6, C.8).

Until their discovery, the existence of a Lagrangian compatible with the self-duality condition
seemed unlikely and its structure appeared at first complicated. Some efforts were needed to demonstrate
that the Avdeev-Chizhov tensors describe a complex scalar field with one real degrees of freedom for $B$ and one for $\BB$,
and to delineate their possible interactions \cite{LRS95,Wet08}.
With hindsight, we can reconstruct the model just from the rules of quantum field theory.
The possible couplings of a 2-tensor to a chiral Fermion are strongly constrained by Lorentz invariance.
The $\mu\nu$ indices must act on the Fermions via the antisymmetrized product of two
Pauli matrices (see appendix B for our precise notations) and this product is by itself self-dual:
\BE
\SG\SB =  P^+\;\SG\SB\;,\;\;\SB\SG =  P^-\;\SB\SG\;,\;\;
\EE
where $P^{\pm}$ are the self-duality projectors
\BQ
P^{\pm}_{\mu\nu\rho\sigma} = \frac {1}{4} (g_{\mu\nu}g_{\rho\sigma} - g_{\mu\rho}g_{\nu\sigma} \mp i\,\epsilon_{\mu\nu\rho\sigma})\;,
\\
(P^+)(P^+) = P^+\;,\;\;(P^-)(P^-) = P^-\;,\;\;(P^+)(P^-) = (P^-)(P^+) = 0\;.\;\;
\EQ
Therefore, the only antisymmetric tensors which can couple to chiral Fermions are self or
anti-self-dual.
The anti-self-dual field $B$ absorbs right states and emits left states, and the
self-dual field $\BB$ absorbs left states and emits right states according to the Feynman diagrams:

$\;\;\;\;\;\;\;\;\;\;\;\;\;\;\;$
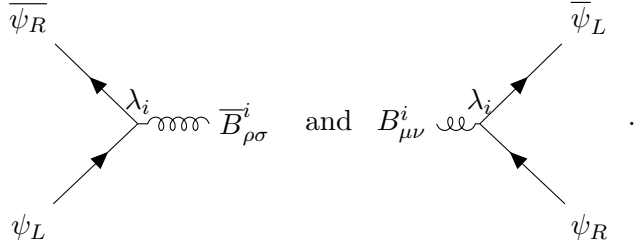
\begin{tikzpicture}
\begin{feynman}
\vertex (a) {\(\BB^i_{\rho\sigma}\)\,\,\,};
\vertex [left = of a, label=\(\lX_i\)] (x);
\vertex [below left=of x] (b){\(\psi_L\)};
\vertex [above left=of x] (c){\(\overline{\psi_R}\)};
\diagram* {
  (a) -- [gluon] (x),
  (x) --  [anti fermion](b),
  (x) --  [fermion](c),
};
\vertex  [right = of a] (a2) {\mbox{and}\,\,\, \(B_{\mu\nu}^i\)};
\vertex [right = of a2, label=\(\lX_i\)] (x2);
\vertex [below right=of x2] (b2){\(\psi_R\)};
\vertex [above right=of x2] (c2){\(\psiB_L\)};
\diagram* {
  (a2) -- [gluon] (x2),
  (x2) --  [anti fermion](b2),
  (x2) --  [fermion](c2),
};
\end{feynman}
\draw[thick] ($(x2) + (2,0)$) node {.};
\end{tikzpicture}

Assuming the standard propagator for the chiral Fermions defined by the Lagrangian
\BQ
\LAG = i \overline {(\psi_R)} \;\SG^{\mu}\partial_{\mu}\;\psi_R
+ i \overline {(\psi_L)} \; \SB^{\mu}\partial_{\mu}\;\psi_L 
\;,
\EQ
the knowledge of these 2 vertices is sufficient to compute the pole part (appendix D) of the propagator of the $\BB B$ field
by closing the Fermion loop:

$\;\;\;\;\;\;\;\;\;\;\;\;\;\;\;\;\;\;\;\;$
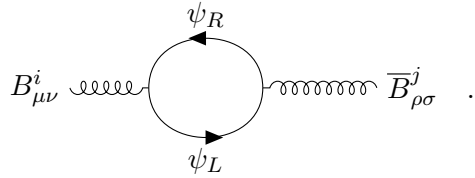
\begin{tikzpicture}
\begin{feynman}
\vertex (a) {\(B^i_{\mu\nu}\)};
\vertex [right=of a] (b);
\vertex [right=of b] (c);
\vertex [right=of c] (d){\(\BB^j_{\rho\sigma}\)\quad \mbox{.}};
\diagram* {
  (a) -- [gluon] (b),
  (b) -- [anti fermion, half left, edge label =\(\psi_R\) ](c),
  (c) -- [anti fermion, half left, edge label =\(\psi_L\) ] (b),
  (d) -- [gluon] (c),
};
\end{feynman}
\end{tikzpicture}

Carefully computing this Feynman diagram (appendix E), we recover the tensorial structure
of the Avdeev-Chizhov propagator \cite{AC94}
\BE
\LAG_B = - \kappa_{ij}g^{\mu\nu} \;\;\partial^{\alpha} \BB^i_{\alpha\mu}\;\;\partial^{\beta} B^j_{\beta\nu}\;\;,
\EE
however \cite{TM20b}, an unexpected consequence of the chiral couplings
of the $\BB$ $B$ fields  (2.10) is that the $\kappa_{ij}$ metric is calculated as a chiral trace:
\BE
\kappa_{ij} = \frac{1}{2}\; Tr ((1+\chi)\lX_i\lX_j) =  \frac{1}{2}\;Tr (\lX_i\lX_j) + \frac{1}{2}\;STr (\lX_i\lX_j)
\;\;.
\EE
The theory hesitates between a Lie algebra-like metric: $Tr(\lX_i\lX_j)$, and a Lie-Kac superalgebra supermetric: $STr(\lX_i\lX_j)$.
The resolution of this dilemma depends on the number and types of chiral Fermions described by the model and is discussed below in section 6.

\section {The Bosonic interaction terms are induced by the Fermion loops}

Following our discussion of the Avdeev-Chizhov fields, we now extend the method to
determine the propagators and self-interactions of the remaining components of the superchiral superconnection.
We postulate
the generalized Dirac Lagrangian 
\BQ
\LAG = i\; \overline {(\psi)} \;\DTS\;\psi\;,
\EQ
where $\DT = \chi d + \AT$, and $\AT$ is our new superchiral superconnection (2.7).
The renormalization of the wave functions upon inclusion of a Fermion loop as above
gives the usual propagator of the scalars and the vectors, as well as the Avdeev-Chizhov propagator \cite{AC94} as derived in (3.6):
\BQ
\LAG_{\Phi} = - \kappa_{ij}\;g^{\mu\nu} \;\;\partial_{\mu} \PhiB^i\;\;\partial_{\nu} \Phi^j\;,\;\;
\\
\LAG_A = - \frac {1}{4}\;\kappa_{ab}g^{\mu\rho}g^{\nu\sigma} \;\;(\partial_{\mu}A^a_{\nu} - \partial_{\nu}A^a_{\mu}) \;\;(\partial_{\rho}A^b_{\sigma} - \partial_{\sigma}A^b_{\rho})\;,
\\
\LAG_{B} = - \kappa_{ij}\;g^{\mu\nu} \;\;\partial^{\alpha} \BB^i_{\alpha\mu}\;\;\partial^{\beta} B^j_{\beta\nu}\;,
\EQ
where the same $\kappa_{ij}$ metric controls the scalar (4.2) and tensor (3.7) propagators.
The vector metric $\kappa_{ab} = g_{ab} = \frac{1}{2}\;Tr (\lX_a\lX_b)$ is the only term that is purely Lie-algebra like and does not
hesitate.

The interaction terms are given by the pole part of the Fermion loops with 3 external fields. The Feynman diagrams

$\;\;\;\;\;\;\;\;\;\;$
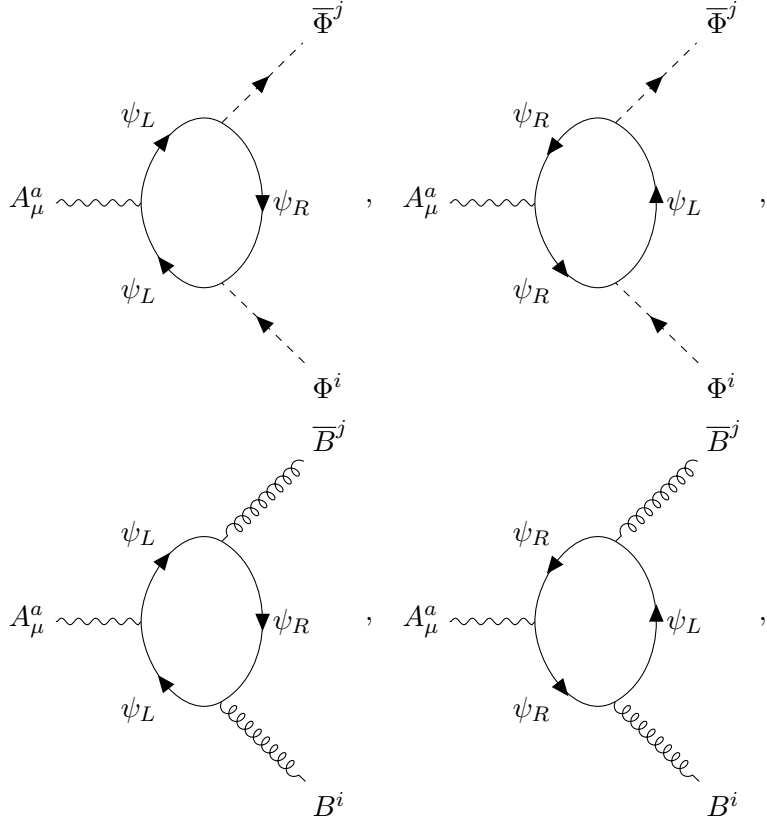
\begin{tikzpicture}
\begin{feynman}
  \vertex (a1) {\(A^a_{\mu}\)};
\vertex [right= of a1] (a);
\vertex [below right=of a] (b);
\vertex [above right=of a] (c);
  \vertex [below right=of b](b1) {\(\Phi^i\)};
  \vertex [above right=of c](c1) {\(\PhiB^j\)};
\diagram* {
  (a1) -- [photon] (a),
  (b1) -- [charged scalar] (b),
  (c) -- [charged scalar] (c1),
  (a) --  [fermion, in=150,out=90, edge label =\(\psi_L\) ](c),
  (c) -- [fermion,  in=30,out=-30, edge label =\(\psi_R\) ] (b),
  (b) -- [fermion, in=270,out=210, edge label =\(\psi_L\) ] (a),
};
\end{feynman}
\draw[thick] ($(a) + (3,0)$) node {,};
\end{tikzpicture}
\begin{tikzpicture}
\begin{feynman}
  \vertex (a1) {\(A^a_{\mu}\)};
\vertex [right= of a1] (a);
\vertex [below right=of a] (b);
\vertex [above right=of a] (c);
  \vertex [below right=of b](b1) {\(\Phi^i\)};
  \vertex [above right=of c](c1) {\(\PhiB^j\)};
\diagram* {
  (a1) -- [photon] (a),
  (b1) -- [charged scalar] (b),
  (c) -- [charged scalar] (c1),
  (a) --  [anti fermion, in=150,out=90, edge label =\(\psi_R\) ](c),
  (c) -- [anti fermion, in=30,out=-30, edge label =\(\psi_L\) ] (b),
  (b) -- [anti fermion, in=270,out=210, edge label =\(\psi_R\) ] (a), 
};
\end{feynman}
\draw[thick] ($(a) + (3,0)$) node {,};
\end{tikzpicture}

$\;\;\;\;\;\;\;\;\;\;$
\begin{tikzpicture}
\begin{feynman}
  \vertex (a1) {\(A^a_{\mu}\)};
\vertex [right= of a1] (a);
\vertex [below right=of a] (b);
\vertex [above right=of a] (c);
  \vertex [below right=of b](b1) {\(B^i\)};
  \vertex [above right=of c](c1) {\(\BB^j\)};
\diagram* {
  (a1) -- [photon] (a),
  (b) -- [gluon] (b1),
  (c1) -- [gluon] (c),
  (a) --  [fermion, in=150,out=90, edge label =\(\psi_L\) ](c),
  (c) -- [fermion,  in=30,out=-30, edge label =\(\psi_R\) ] (b),
  (b) -- [fermion, in=270,out=210, edge label =\(\psi_L\) ] (a),
};
\end{feynman}
\draw[thick] ($(a) + (3,0)$) node {,};
\end{tikzpicture}
\begin{tikzpicture}
\begin{feynman}
  \vertex (a1) {\(A^a_{\mu}\)};
\vertex [right= of a1] (a);
\vertex [below right=of a] (b);
\vertex [above right=of a] (c);
  \vertex [below right=of b](b1) {\(B^i\)};
  \vertex [above right=of c](c1) {\(\BB^j\)};
\diagram* {
  (a1) -- [photon] (a),
  (b) -- [gluon] (b1),
  (c1) -- [gluon] (c),
  (a) --  [anti fermion, in=150,out=90, edge label =\(\psi_R\) ](c),
  (c) -- [anti fermion, in=30,out=-30, edge label =\(\psi_L\) ] (b),
  (b) -- [anti fermion, in=270,out=210, edge label =\(\psi_R\) ] (a), 
};
\end{feynman}
\draw[thick] ($(a) + (3,0)$) node {,};
\end{tikzpicture}

induce the expected covariant derivative minimal coupling
\BE
\LAG = - D_{\mu}\PhiB\;D_{\mu}\Phi - D^{\alpha}\BB_{\alpha\mu}\;D^{\beta}B_{\beta\mu}\;.
\EE
But there is a caveat \cite{TM20b}: since the orientation of the loop is correlated with the chirality of the looping Fermions,
the interaction term hidden in the definition of the covariant derivative
\BQ
D_{\mu}\Phi_i = \partial_{\mu}\Phi_i + t_{aij} A^a_{\mu}\Phi^j\;,
\EQ
\BQ
D^{\alpha}B_{i\alpha\mu} = \partial^{\alpha}B_{i\alpha\mu} + t_{aij} A^{a\alpha}B^j_{\alpha\mu}\;,
\EQ
is given by the chiral trace
\BE
t_{aij} = Tr ((1+\chi) \;\lX_a\lX_i\lX_j - (1-\chi) \;\lX_a\lX_j\lX_j) = Tr (\lX_a\;[\lX_i,\lX_j]) + STr (\lX_a\;\{\lX_i,\lX_j\})\;.
\EE
As found for the tensor propagators (3.7 and 4.2), the $t_{aij}$
interaction terms (4.6) are neither fish nor meat.
They hesitate between a Lie algebra trace and a Lie-Kac superalgebra supertrace.
They are not universal. They depend on the Fermion content of the model.

Another novelty is the apparition of a new
mixed $AB\PhiB$ coupling, which must be considered as a genuine component of the superchiral minimal
coupling, induced by the Feynman diagrams:

$\;\;\;\;\;\;\;\;\;\;$
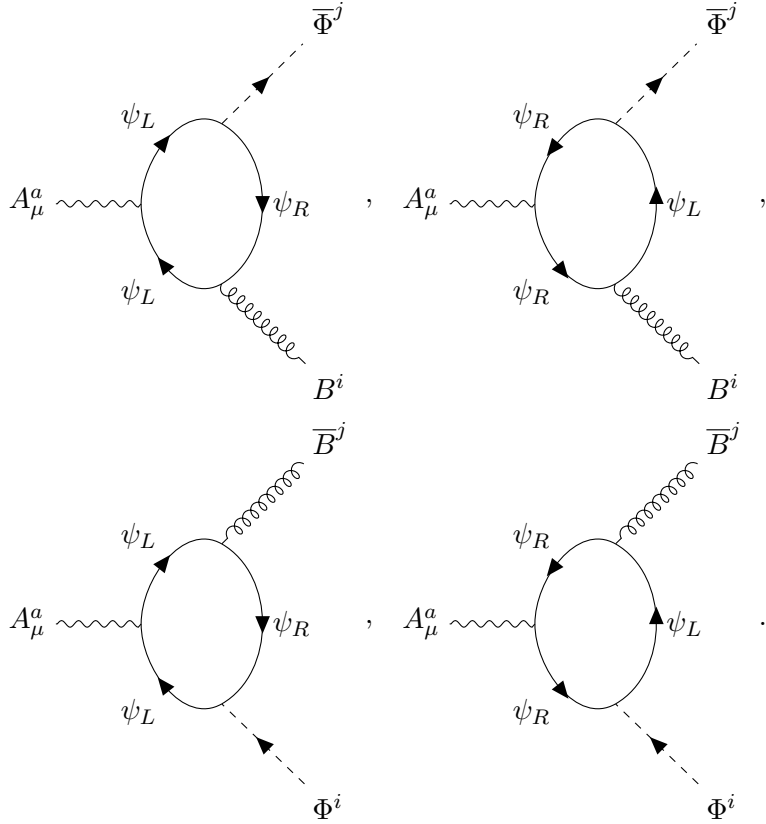
\begin{tikzpicture}
\begin{feynman}
  \vertex (a1) {\(A^a_{\mu}\)};
\vertex [right= of a1] (a);
\vertex [below right=of a] (b);
\vertex [above right=of a] (c);
  \vertex [below right=of b](b1) {\(B^i\)};
  \vertex [above right=of c](c1) {\(\PhiB^j\)};
\diagram* {
  (a1) -- [photon] (a),
  (b) -- [gluon] (b1),
  (c) -- [charged scalar] (c1),
  (a) --  [fermion, in=150,out=90, edge label =\(\psi_L\) ](c),
  (c) -- [fermion,  in=30,out=-30, edge label =\(\psi_R\) ] (b),
  (b) -- [fermion, in=270,out=210, edge label =\(\psi_L\) ] (a),
};
\end{feynman}]
\draw[thick] ($(a) + (3,0)$) node {,};
\end{tikzpicture}
\begin{tikzpicture}
\begin{feynman}
  \vertex (a1) {\(A^a_{\mu}\)};
\vertex [right= of a1] (a);
\vertex [below right=of a] (b);
\vertex [above right=of a] (c);
  \vertex [below right=of b](b1) {\(B^i\)};
  \vertex [above right=of c](c1)  {\(\PhiB^j\)};
\diagram* {
  (a1) -- [photon] (a),
  (b) -- [gluon] (b1),
  (c) -- [charged scalar] (c1),
  (a) --  [anti fermion, in=150,out=90, edge label =\(\psi_R\) ](c),
  (c) -- [anti fermion, in=30,out=-30, edge label =\(\psi_L\) ] (b),
  (b) -- [anti fermion, in=270,out=210, edge label =\(\psi_R\) ] (a), 
};
\end{feynman}
\draw[thick] ($(a) + (3,0)$) node {,};
\end{tikzpicture}

$\;\;\;\;\;\;\;\;\;\;$
\begin{tikzpicture}
\begin{feynman}
  \vertex (a1) {\(A^a_{\mu}\)};
\vertex [right= of a1] (a);
\vertex [below right=of a] (b);
\vertex [above right=of a] (c);
  \vertex  [below right=of b](b1)  {\(\Phi^i\)};
  \vertex [above right=of c](c1){\(\BB^j\)};
\diagram* {
  (a1) -- [photon] (a),
  (b1) -- [charged scalar] (b),
  (c1) -- [gluon] (c),
  (a) --  [fermion, in=150,out=90, edge label =\(\psi_L\) ](c),
  (c) -- [fermion,  in=30,out=-30, edge label =\(\psi_R\) ] (b),
  (b) -- [fermion, in=270,out=210, edge label =\(\psi_L\) ] (a),
};
\end{feynman}
\draw[thick] ($(a) + (3,0)$) node {,};
\end{tikzpicture}
\begin{tikzpicture}
\begin{feynman}
  \vertex (a1) {\(A^a_{\mu}\)};
\vertex [right= of a1] (a);
\vertex [below right=of a] (b);
\vertex [above right=of a] (c);
  \vertex [below right=of b](b1) {\(\Phi^i\)};
  \vertex [above right=of c](c1) {\(\BB^j\)};
\diagram* {
  (a1) -- [photon] (a),
  (b1) -- [charged scalar] (b),
  (c1) -- [gluon] (c),
  (a) --  [anti fermion, in=150,out=90, edge label =\(\psi_R\) ](c),
  (c) -- [anti fermion, in=30,out=-30, edge label =\(\psi_L\) ] (b),
  (b) -- [anti fermion, in=270,out=210, edge label =\(\psi_R\) ] (a), 
};
\end{feynman}
\draw[thick] ($(a) + (3,0)$) node {.};
\end{tikzpicture}

The tensorial structure of these counterterms is unusual because the propagator (3.6) of the $\BB B$ field
has a complicated structure
\BE
P^+_{\mu\nu\alpha\beta}\;k^{\alpha} g^{\beta\gamma}k^{\delta}\;P^-_{\gamma\delta\rho\sigma}/(k^2)^2\;.
\EE
When we perform the calculation, we get with the same strength as in (4.3) the interaction:
\BE
\LAG_{AB\Phi} = \frac{1}{4}\;t_{aij} \;F^a_{\mu\nu} (\BB^i_{\mu\nu}\Phi^j + B^i_{\mu\nu}\PhiB^j)\;.
\EE
This is the only term which is Lorentz invariant and invariant under the Lie subalgebra.
The coupling matrix $t_{aij}$ is the same mixture (4.6) of trace and supertrace that appeared above
in $D\Phi$ and $DB$, and is common to $A\PhiB\Phi$, $A\BB B$, $A\BB\Phi$ and $AB\PhiB$ 
because the $\Phi$ and the $B$ fields have the same chiral interactions with the Fermions (2.9,2.10). Regrouping all terms we get
\BE
\LAG_{B\Phi} = - \kappa_{ij} \;\;D^{\alpha} \BB^i_{\alpha\mu}\;\;D^{\beta} B^j_{\beta\mu}\;\; -
             \kappa_{ij} \;\;D^{\alpha} \PhiB^i\;\;D_{\alpha} \PhiB^j\;\; -
             \frac{1}{4}\,t_{aij} \;F^{a\mu\nu} (\BB^i_{\mu\nu}\Phi^j + B^i_{\mu\nu}\PhiB^j)\;.
\EE
The interesting point is that the $F$ coupling cannot be freely adjusted.
It comes as a consequence of the $\DTS$ coupling of all
the connection fields to the Fermions and should be considered as an indispensable part of the
minimal coupling of the Avdeev-Chizhov fields. The same couplings
appear in (2.11) as part of the classic Lagrangian $\FHS^2$.

\section {The Adler-Bell-Jackiw vector anomaly viewed as superalgebraic}

The main surprise of the previous calculations is that the theory seems to hesitate between
a Lie algebra and a Lie superalgebra structure. The scalar propagator $\kappa_{ij}$ (3.7) and the
vector-scalar or vector-tensor vertex $t_{aij}$ (4.6) contain a Lie algebra and a Lie superalgebra
tensor, which cannot both be well defined at the same time. But \textit{a posteriori}, this is not
so surprising; this situation is actually very well known in physics. If we compute just as
before the chiral Fermion loop contributions to the triple vector interaction:

$\;\;\;\;\;\;\;\;\;\;$
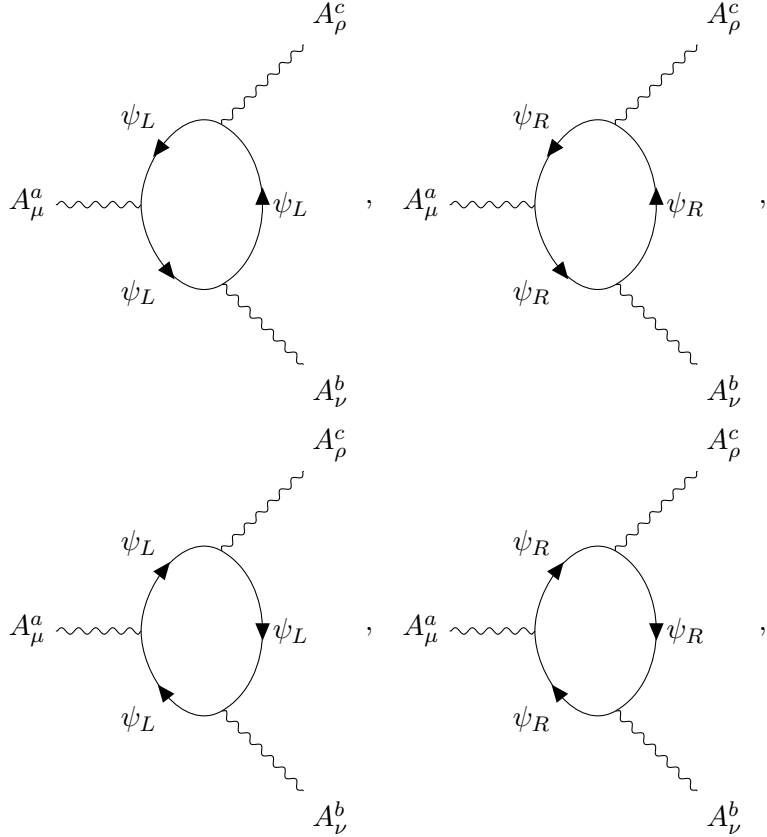
\begin{tikzpicture}
\begin{feynman}
  \vertex (a1) {\(A^a_{\mu}\)};
\vertex [right= of a1] (a);
\vertex [below right=of a] (b);
\vertex [above right=of a] (c);
  \vertex [below right=of b](b1) {\(A^b_{\nu}\)};
  \vertex [above right=of c](c1) {\(A^c_{\rho}\)};
\diagram* {
  (a1) -- [photon] (a),
  (b1) -- [photon] (b),
  (c1) -- [photon] (c),
  (a) --  [anti fermion, in=150,out=90, edge label =\(\psi_L\) ](c),
  (c) -- [anti fermion, in=30,out=-30, edge label =\(\psi_L\) ] (b),
  (b) -- [anti fermion, in=270,out=210, edge label =\(\psi_L\) ] (a), 
};
\end{feynman}
\draw[thick] ($(a) + (3,0)$) node {,};
\end{tikzpicture}
\begin{tikzpicture}
\begin{feynman}
  \vertex (a1) {\(A^a_{\mu}\)};
\vertex [right= of a1] (a);
\vertex [below right=of a] (b);
\vertex [above right=of a] (c);
  \vertex [below right=of b](b1) {\(A^b_{\nu}\)};
  \vertex [above right=of c](c1) {\(A^c_{\rho}\)};
\diagram* {
  (a1) -- [photon] (a),
  (b1) -- [photon] (b),
  (c1) -- [photon] (c),
  (a) --  [anti fermion, in=150,out=90, edge label =\(\psi_R\) ](c),
  (c) -- [anti fermion, in=30,out=-30, edge label =\(\psi_R\) ] (b),
  (b) -- [anti fermion, in=270,out=210, edge label =\(\psi_R\) ] (a), 
};
\end{feynman}
\draw[thick] ($(a) + (3,0)$) node {,};
\end{tikzpicture}

$\;\;\;\;\;\;\;\;\;\;$
\begin{tikzpicture}
\begin{feynman}
  \vertex (a1) {\(A^a_{\mu}\)};
\vertex [right= of a1] (a);
\vertex [below right=of a] (b);
\vertex [above right=of a] (c);
  \vertex [below right=of b](b1) {\(A^b_{\nu}\)};
  \vertex [above right=of c](c1) {\(A^c_{\rho}\)};
\diagram* {
  (a1) -- [photon] (a),
  (b1) -- [photon] (b),
  (c1) -- [photon] (c),
  (a) --  [fermion, in=150,out=90, edge label =\(\psi_L\) ](c),
  (c) -- [fermion,  in=30,out=-30, edge label =\(\psi_L\) ] (b),
  (b) -- [fermion, in=270,out=210, edge label =\(\psi_L\) ] (a),
};
\end{feynman}
\draw[thick] ($(a) + (3,0)$) node {,};
\end{tikzpicture}
\begin{tikzpicture}
\begin{feynman}
  \vertex (a1) {\(A^a_{\mu}\)};
\vertex [right= of a1] (a);
\vertex [below right=of a] (b);
\vertex [above right=of a] (c);
  \vertex [below right=of b](b1) {\(A^b_{\nu}\)};
  \vertex [above right=of c](c1) {\(A^c_{\rho}\)};
\diagram* {
  (a1) -- [photon] (a),
  (b1) -- [photon] (b),
  (c1) -- [photon] (c),
  (a) --  [fermion, in=150,out=90, edge label =\(\psi_R\) ](c),
  (c) -- [fermion,  in=30,out=-30, edge label =\(\psi_R\) ] (b),
  (b) -- [fermion, in=270,out=210, edge label =\(\psi_R\) ] (a),
};
\end{feynman}
\draw[thick] ($(a) + (3,0)$) node {,};
\end{tikzpicture}

we also obtain 2 types of terms proportional to:
\BQ
Z_f = Tr(\lX_a\;[\lX_b,\lX_c])\;\;\;A^{a \mu}A^{b \nu}\partial_{\mu}A^c_{\nu}
\;,\\
d_{abc} = STr(\lX_a\;\{\lX_b,\lX_c\})
\;.
\EQ
The $Tr(\lX_a\;[\lX_b,\lX_c])$ term is the expected
counterterm to the Lie algebra triple vector vertex
contained in the classical Yang-Mills Lagrangian $Tr(F^2)$.
The $d_{abc} = STr(\lX_a\;\{\lX_b,\lX_c\})$ term  is the surprise Adler-Bell-Jackiw
anomaly \cite{Adler,BJ} coming from the measure of the chiral Fermions (see for example section 5 of the lectures of Bilal \cite{Bilal}),
where the supertrace is defined
as the trace over the left Fermions minus the trace over the right Fermions..
Using the superchirality condition (1.1), we can reinterpret this helicity supertrace in the sense of
Herman Weyl (B.1), as the internal superalgebra supertrace in the sense of Kac (1.2)
and (A.1), and identify the vector anomaly with the even part of the rank-3 super-Casimir
operator (A.7) of the superalgebra. The role of the Hodge dual in the Adler anomaly is
also consistent with our superchiral condition (1.1) which correlates the chirality of the
spinor bundle with the Hodge duality of the exterior bundle (1.3).

To conclude, the triple-vector vertex (5.1) also hesitates between a Lie algebra and a
Lie superalgebra structure. The Adler-Bell-Jackiw anomalous term (5.1) is superalgebraic
in nature and cancels out if the supertrace of the Casimir of rank 3 (A.7) of the Lie
superalgebra vanishes.

The Fermion loop counterterms to the quartic vertices $A^4$, $A^2\PhiB\Phi$, $A^2\BB B$, $A^2\BB \Phi$, $A^2\PhiB B$
also contain anomalies, but they automatically follow the structure of the cubic
terms because of the Lie algebra Ward identities. For example the classic $A^4$ vertices are
the complements of the $A^3$ vector terms in the classic Yang-Mills Lagrangian $Tr(F^2)$. The
$(A^2\BB\Phi)$ counterterm is the complement of the $(A\BB \Phi)$ term in the $(F\{\BB,\Phi\})$ Lagrangian.
The quartic potentials $\PhiB^2 \Phi^2$ , $(\PhiB\Phi\BB B)$ and $(\BB^2 B^2)$ remain to be studied.

\section {Classification of the anomaly-free superchiral superconnections}

We have identified three obstructions to the construction of the quantum field theory: (3.7), (4.6) and (5.1).
We wish to show here that these hesitations between trace and supertrace are resolved in many superchiral models.

Consider first the scalar anomalies.
Since the trace operator is invariant under circular permutation, we can use the
closure relation (A.3) of the superalgebra to rewrite the trace term in (3.7) as
\BE
Tr (\lX_i\lX_j) = Tr (\lX_j\lX_i) = \frac{1}{2}Tr (\{\lX_i,\lX_j\}) = \frac{1}{2} d^a_{ij} Tr (\lX_a)\;.
\EE
In the same way, we can rewrite (4.6) as
\BQ
Tr (\lX_a\;[\lX_i,\lX_j]) = Tr (\lX_a\lX_i\lX_j  -\lX_a\lX_j\lX_i) \\
= Tr (\lX_i\lX_j\lX_a  -\lX_i\lX_a\lX_j)
= - Tr (\lX_i\;[\lX_a,\lX_j]) \\
= - f^k_{aj} Tr (\lX_i\lX_k) = - \frac{1}{2} \; d^b_{ik}f^k_{aj}\;Tr(\lX_b)\;.
\EQ
Hence if all the even generators satisfy the constraint
\BE
  Tr (\lX_a) = 0\;,
\EE
the theory is superalgebraic: the propagators of the scalars and of the Avdeev-Chizhov tensors are
controlled by the odd part of the super-Killing metric $\kappa_{ij} =\frac{1}{2} STr(\lX_i\lX_j)$\,,
and their interactions with the vectors
are governed by the symmetric structure constants of the superalgebra $t_{aij} = d_{aij}$. 
For the $U(1)$ factor, this constraint is non trivial.

Consider now the vector anomaly (5.1).
The simple Lie algebras are of type $A$, $B$, $C$, $D$, $E$ and $F$. Among those, only $A_m = SU(m+1), \;m \ge 2$, including 
$A_3 = D_3 = SO(6) = SU(4)$\,,
admit a Casimir of rank 3. In addition, we can have a $U(1)$ algebra, denoted $Y$, which
generates two supplementary Casimirs of rank 3: $Y^3$ and $Y C_2$ where $C_2$ is a rank 2 Casimir
of any other Lie algebra present in the model.
A superchiral model associated to the simple Kac superalgebras $G(3)$ or $OSp(m/n)$ with $m \ge 7$
has no Casimir of rank 3 and no $U(1)$ factor,
so it cannot have a vector anomaly. As all its generators are traceless (6.3),
it has no scalar anomaly either. Therefore the model is superalgebraic and anomaly-free.
It is nevertheless chiral whenever the
non-Abelian charges of the left and right Fermions differ.

\section {Anomaly cancellation in the superchiral SU(2/1) model of leptons and quarks}

It is also possible to cancel the scalar anomalies by combining several irreducible representations.
For example, in the $SU(2/1)$ model of the electroweak interactions,
the superchirality condition (1.1) implies that $SU(2)$ only acts on the left doublets (2.8).
The cumulative hypercharge of the \{right-electron/(left-electron,left-neutrino)\} triplet
$Tr(Y) = -4$ (\cite{TM20b} appendix B) is compensated by the cumulative hypercharge of the 3 colored
(up,down) quarks quadruplets 
$Tr(Y) = 3 \times 4/3$ (\cite{TM20b} appendix D).
This is equivalent to the observation that the electric charge of the hydrogen atom
(1 electron plus three $uud$ quarks) vanishes. Using (6.1) and (6.2), the scalar anomalies (3.7) and (4.6) cancel out.
As found in 1972 by Bouchiat, Iliopoulos and Meyer (BIM \cite{BIM}), the four vector anomalies
$Y^3$, $Y\;SU(2)^2$, $Y\; SU(3)^2$  and $SU(3)^3$ (5.2) also cancel out in the standard model, separately for each family.
Indeed, the complete rank 3 super-Casimir tensor cancels for each family, removing any potential measure anomaly
$STr(\lX_a\;[\lX_i,\lX_j])$ in the $A\BB B$ and $A \BB \Phi$ triangle diagrams.

\section {Discussion}

The concept of a superconnection defined as an odd polyform, a linear combination
of exterior-forms of all degrees valued in a Lie superalgebra (2.2),
was first introduced by Thierry-Mieg and Ne'eman in 1982 \cite{NTM82} in terms of the primitive forms $(\phi,a,b,c,e)$ as
\BE
\AT = (\phi + b + e)^i\lX_i + (a + c)^a\lX_a\;,
\EE
and then by Quillen and Mathai in their seminal papers \cite{Quillen,MQ86} and
recently modified in \cite{TM20a}. 
In Quillen \cite{Quillen,MQ86},
the covariant differential is defined as $\DT = d + A + L$, where $L = L^i\lX_i$ is, as for us, a mixed exterior-form of even
degree valued in the odd module of the superalgebra. But because $L$ must be odd relative to the
differential calculus to ensure that the curvature $\FT = \DT\DT$ defines a linear map,
Quillen assumes that the components $L^i$ of $L$ are valued in another graded algebra
which anticommutes with the exterior-forms. The difficulty
is that these partially anticommuting $L^i$ cannot be represented in quantum field theory by commuting scalar fields.
This is probably why the works of Ne'eman and Sternberg \cite {NSF} or of the Marseille-Mainz
group (see for example \cite{CQ0,HS}) who have all adopted the Quillen formalism,
stop short of the quantum theory.
In our construction \cite{TM20a}, the components $\phi^i$ of $\phi = \phi^i\lX_i$ are
just ordinary commuting functions. Nevertheless $\phi$ is odd with respect to our differential calculus,
as requested by Quillen \cite{Quillen}, because the $\lX_i$ matrices anticommute with the chirality $\chi$ (2.1)
which decorates our exterior differential $\dT = \chi d$ (2.2). As a result, the commuting $\phi^i$ can
be represented by Bose scalars and
we can develop a quantum field theory formalism as here. This modifies the calculation
of the superconnection cohomology \cite{MQ86} which should be reexamined and we conjecture
that the Adler-Bell-Jackiw quantum anomalies play a role as obstructions in this purely
geometrical context. However, an inconvenience of the superconnection formalism is the
excessive number of component fields

The new self-dual superchiral constraints $\AT = ^*\AT\; \chi$ introduced in the present work as
equation (1.1) provides a subtle way to reduce this number. The Hodge duality conspires with the chiral properties of the Pauli matrices
such that the action of the Dirac operator is focused on a single chirality (2.6).
If $\AT$ is self-dual, it only absorbs left Fermions; if it is anti-self-dual, it only absorbs right Fermions. Coupling the Hodge duality signs (C.1) with the superalgebra supertrace operator $\chi$,
defined as $Str(\cdots) = Tr (\chi \cdots)$ ((1.2) and appendix A), links the spinor bundle (appendix B) to the
exterior bundle (appendix C) and focuses the action of the superconnection (2.8)--(2.10)
on the Landau charge-parity $CP$ positive Fermions: $\ATS = \ATS \;(1 \pm \chi)(1 \mp \gamma_5)/4$. The constraint also
eliminates the primitive fields (2.2) in favor of the self-chiral fields $(\Phi,A,B)$ (2.8-10).
Applied to the $SU(2/1)$ superalgebraic model of the elementary particles, we recover
the vector interactions postulated \textit{ex nihilo} by Ne'eman \cite{N1} and Fairlie \cite{F1} and the scalar interactions
postulated by Thierry-Mieg in \cite{TM20b}. The superchirality focuses, as observed in Nature,
the action of the $SU(2)$ vector Bosons on
the left leptons and quarks (section 7).

Because the left and right leptons and quarks do not carry the same $SU(2)$ or $U(1)$ charges, the Yang-Mills
sector is subject to the Adler-Bell-Jackiw anomaly \cite {Adler,BJ} which curiously (section 5) has a superalgebraic
symmetric structure  $C_{abc} = STr (\lX_a\;\{\lX_b,\lX_c\})$ involving three Lie algebra even indices $(abc)$.
Our new interpretation of this well-known result is that
the chiral supertrace, in the sense of Adler-Bell-Jackiw: left minus right Fermions, is
equivalent to the charge supertrace in the sense of Kac (1.2), so the supertrace terms in (5.2)
matches the definition of the even part of the super-Casimir (A.7).
In a similar way, the counterterms to the scalar and tensor interactions (4.6) involve algebraic
antisymmetric structure constants $f_{aij} = Tr (\lX_a\;[\lX_i,\lX_j])$ although $(ij)$ are odd indices.
We find that these unwanted terms cancel out if all the even generators are traceless (6.3).
This is true in particular
in the standard model of the fundamental interactions when we apply the BIM \cite{BIM}
mechanism whereby each lepton family is balanced by its pair of quarks (section 7).
The complete rank-3 super-Casimir tensor (A.7) vanishes and only the representation independent universal couplings
$f_{abc} = Tr (\lX_a\;[\lX_b,\lX_c])$
and $d_{aij} = STr (\lX_a\;\{\lX_i,\lX_j\})$ 
survive. The resulting scalar-vector-tensor
theory is therefore, at one-loop, superalgebraic and anomaly-free.

Another interesting consequence of our superchiral structure is the
induction by the Fermion loops of a new scalar-vector-tensor
triple interaction (4.8) which reproduces, if and only if we apply the BIM mechanism,
the structure of the square of the geometric supercurvature (2.11). Once again,
Differential Geometry and Quantum Field Theory agree, conditional on the elimination
of the Adler-Bell-Jackiw anomaly.

These results are tantalizing from a theoretical point of view, yet very surprising: the coupling (2.8-10)
of the vectors $\chi \lX_a$ and of the scalars
$\lX_i (1 \pm \chi)$ are outside the naive superalgebra generated by the $(\lX_a, \lX_i)$ matrices
(appendix A); the couplings to the Fermions are all even (they transform Fermions into
Fermions, not Fermions into Bosons as in Wess-Zumino supersymmetry);  yet the signs induced by the helicity of the
Fermion propagators
restore the superalgebraic structure (4.6), if and only if the model is anomaly-free.
The expected minimal coupling of the vectors to the scalars and the tensors, via the covariant derivatives, is
also necessarily completed by a new scalar-vector-tensor vertex (4.8) which modifies
the asymptotic behavior of the coupling of the scalars and tensors to the Fermions.
A deeper understanding of these equations must be possible.

These results are also curious from a phenomenological point of view even if the
superalgebraic structure is a consequence of the
experimentally verified BIM mechanism whereby the chiral quantum anomalies
are canceled by the balances of the leptons against the quarks.
However the model is highly constrained and offers no choice.
The field content is defined by the differential geometry,
the dynamics are induced by the Fermion loops and there are no
free parameter, except that the Cabbibo-Kobayashi-Maskawa angles
can be understood as specifying the 
details of the 3-generations indecomposable representations of $SU(2/1)$ (see \cite{CQ0,HS} and \cite{TM20b}, appendix H).

Many problems remain.
Having established the self-interactions of the Boson fields (4.9), one still needs
to examine if the theory is renormalizable and in particular if the counterterms involving Boson loops
have the correct Lorentz structure, which seems likely, and the correct algebraic structure, which is non-trivial
as we only have Lie-algebra Ward identities.
The scalar potential has to be evaluated. The symmetry breaking pattern of the model must be studied.
Finally, the crucial open question is the eventual existence of a symmetry associated to
the odd-generators of the superalgebra.

\section*{Acknowledgments}
This research was supported by the Intramural Research Program of the National Library of Medicine, National Institute of Health.
We are grateful to Danielle Thierry-Mieg for clarifying the presentation and to Andre Neveu for stimulating discussions.

\appendix

\section {Definition of a chiral superalgebra}\label{A1}

For completeness, this section is copied from appendix A of \cite{TM20b}.

Let us define, using the notations of \cite{TM20b},  a chiral-superalgebra as a finite dimensional
basic classical Lie-Kac superalgebra \cite{Kac1}, graded by chirality.
For example, we could take a superalgebra of type $SU(m/n)$,
or $OSp(m/2n)$ or a product of Lie algebras and superalgebras
like the $SU(2/1)SU(3)$ superalgebra of the standard model.

The superalgebra acts on a finite dimensional space
of massless Fermion states graded by their helicity.
The chirality matrix $\chi$ is diagonal, with eigenvalue $1$ on the
left Fermions and $-1$ on the right Fermions. It defines the supertrace
\BE
STr(\cdots) = Tr (\chi\;\cdots)\;.
\EE
Each generator is represented by 
a finite dimensional matrix of complex numbers (we do not need anticommuting Grassmann numbers).
The even generators are denoted $\lX_a$ and the odd generators $\lX_i$.
$\chi$ commutes with the $\lX_a$ and anticommutes with the $\lX_i$
\BE
 [\chi,\;\lX_a]_- = \{\chi,\;\lX_i\}_+ = 0 \;.
\EE
The $\lX$ matrices close under (anti)-commutation
\BE
  [\lX_a,\;\lX_b]_- = f^c_{ab} \;\lX_c
\;,\;\;\;
  [\lX_a,\;\lX_i]_- = f^j_{ai} \;\lX_j
\;,\;\;\;
  \{\lX_i,\;\lX_j\}_+ = d^a_{ij} \;\lX_a
\;,
\EE
and satisfy the super-Jacobi relation with 3 cyclic permuted terms:
\BE
   (-1)^{AC} \{\lX_A, \{\lX_B,\;\lX_C]] + 
   (-1)^{BA} \{\lX_B, \{\lX_C,\;\lX_A]] + 
   (-1)^{CB} \{\lX_C, \{\lX_A,\;\lX_B]]  = 0\;.
\EE
The quadratic Casimir tensor $(g_{ab},g_{ij})$, also called the super-Killing metric, is defined as
\BQ
  g_{ab} = \frac{1}{2} STr (\lX_a\lX_b)\;,
\\
  g_{ij} = \frac{1}{2} STr (\lX_i\lX_j)\;.
\EQ
The even part $g_{ab}$ of the metric is as usual symmetric,
but because the odd generators anticommute (A.2) with
the chirality hidden in the supertrace (A.1),
its odd part $g_{ij}$ is antisymmetric. 
The structure constants can be recovered from the
supertrace of a product of 3 matrices
\BQ
f_{abc} = g_{ae}\,f^e_{bc} = \frac{1}{2} STr (\lX_a\,[\lX_b,\lX_c]_-)\;,
\\
d_{aij} = g_{ae}\,d^e_{ij} = \frac{1}{2} STr (\lX_a\,\{\lX_i,\lX_j\}_+)\;.
\EQ
The rank 3 Casimir tensor can be defined as :
\BQ
  C_{abc} = \frac{1}{2} STr (\lX_a\;\{\lX_b,\;\lX_c\}_+)\;,
\\
  C_{aij} = \frac{1}{2} STr (\lX_a\;[\lX_i,\;\lX_j]_-)\;.
\EQ
The Casimirs use the ``wrong'' type of commutator;
otherwise, using equation (A.3), the components of (A.7) could each be simplified.
We also have $g_{ai} = 0$ as well as $C_{abi} = C_{ijk} = 0$\,, since the diagonal elements of the product of an odd number of odd matrices necessarily vanish. 
Using these tensors, we can
construct the quadratic and cubic super-Casimir operators
\BE
 K_2 = g^{AB}\;\lX_A\lX_B\;,\;\;\; K_3 = C^{ABC}\;\lX_A\lX_B\lX_C\;,
\EE
where the upper index metric $g^{AB}$ is the inverse of the lower metric $g_{AB}$,
summation over the repeated indices is implied and ranges over
even and odd values $A,B = a,b,\cdots; i,j,\cdots$, and the indices of $C^{ABC}$
are raised using $g^{AB}$.
The Casimir operators $K_2$ and $K_3$ commute with all the generators
of the superalgebra. In an
irreducible representation, they are represented by a multiple of the
identity matrix. In $SU(2/1)$, which has rank 2, they form a basis of
its enveloping superalgebra.

\section{Pauli matrices} \label{A2}

Because $SO(4)$ is isomorphic to $SU(2)+SU(2)$, the $SO(1,3)$ spinors split into left and right
doublets. The projectors are traditionally represented as
\BQ
P_L = \frac{1}{2} (1 - \gamma_5)\;,\;\;   P_R = \frac{1}{2} (1 + \gamma_5)
\EQ
and we have
\BQ
P_L + P_R = 1\;,\;\;P_L P_R = P_R P_L = 0\;,\;\; P_LP_L = P_L\;,\;\;P_RP_R = P_R\;.
\EQ
The Pauli matrices $\SG$ map the right spinors onto the left spinors 
and the $\SB$ matrices map the left spinors onto the right spinors.
\BQ
\SG = P_L\;\SG\;P_R\;,\;\;\SB = P_R\;\SB\;P_L\;.
\EQ
In Minkowski space they can be represented in $2\times 2$ block form, with non vanishing entries
\BQ
\SG_0 = \begin{pmatrix} 
  \;1 & \;0 \cr \;0 & \;1
\end{pmatrix}
\;,\;\;\;  \SG_1 = \begin{pmatrix} 
  \;0 & \;1 \cr \;1 & \;0
\end{pmatrix}
\;,\;\;\;  \SG_2 = \begin{pmatrix} 
 \;0 & -i \cr \;i & \;0
\end{pmatrix}
\;,\;\;\;  \SG_3 = \begin{pmatrix} 
 \;1 & \;0 \cr \;0 & -1
\end{pmatrix} \,,
\\
\\
  \SB_0 = \begin{pmatrix} 
 -1 & \;0 \cr \;0 & -1
\end{pmatrix}
\;,\;\;\;  \SB_1 = \begin{pmatrix} 
 \;0 & \;1 \cr \;1 & \;0
\end{pmatrix}
\;,\;\;\;  \SB_2 = \begin{pmatrix} 
 \;0 & -i \cr \;i & \;0
\end{pmatrix}
\;,\;\;\;  \SB_3 = \begin{pmatrix} 
 \;1 & \;0 \cr \;0 & -1
\end{pmatrix}\,,
\EQ
They are Hermitian, and satisfy the chiral Clifford-Weyl relations
\BQ
\SG_{\mu}\SB_{\nu} + \SG_{\nu}\SB_{\mu} = 2 g_{\mu\nu} \;P_L\;,
\\
\SB_{\mu}\SG_{\nu} + \SB_{\nu}\SG_{\mu} = 2 g_{\mu\nu} \;P_R\;,
\EQ
where $g_{\mu\nu} = g^{\mu\nu}$ denotes the diagonal Minkowski metric $(-1,1,1,1)$. Importantly, if we compute the
trace of the product of four $\SG$ matrices we find a tensor with mixed symmetry
\BQ
  Tr (\SG_{\mu}\SB_{\nu}\SG_{\rho}\SB_{\sigma}) = 2 (g_{\mu\nu}g_{\rho\sigma}\;-\;g_{\mu\rho}g_{\nu\sigma}\;+\;g_{\mu\sigma}g_{\nu\rho}
  +  i\;\epsilon_{\mu\nu\rho\sigma})\;,
\\
  Tr (\SB_{\mu}\SG_{\nu}\SB_{\rho}\SG_{\sigma}) = 2 (g_{\mu\nu}g_{\rho\sigma}\;-\;g_{\mu\rho}g_{\nu\sigma}\;+\;g_{\mu\sigma}g_{\nu\rho}
  - i\;\epsilon_{\mu\nu\rho\sigma})\;,
\EQ
where the $g$ terms are symmetric, and $\epsilon$ is fully antisymmetric in $\mu\nu\rho\sigma$ with $\epsilon_{0123} = 1$,
hence $\epsilon^{0123} = -1$ and $\epsilon_{\mu\nu\rho\sigma}\epsilon^{\mu\nu\rho\sigma} = -24$.
If we now consider the antisymmetric products of several Pauli matrices, we can derive by inspection several
interesting identities:
\BQ
i\;\epsilon_{\mu\nu\rho\sigma}\;\SB^{\rho}\SG^{\sigma}  = \SB_{\mu}\SG_{\nu} - \SB_{\nu}\SG_{\mu}\;,\;\;
i\;\epsilon_{\mu\nu\rho\sigma}\;\SG^{\rho}\SB^{\sigma}  = - \SG_{\mu}\SB_{\nu} + \SG_{\nu}\SB_{\mu}\;,
\\
i\;\epsilon_{\mu\nu\rho\sigma}\;\SG^{\nu}\SB^{\rho}\SG^{\sigma}  = - 6 \;\SG_{\mu}\;,\;\;
i\;\epsilon_{\mu\nu\rho\sigma}\;\SB^{\nu}\SG^{\rho}\SB^{\sigma}  = 6 \;\SB_{\mu}\;,
\\
i\;\epsilon_{\mu\nu\rho\sigma}\;\SG^{\mu}\SB^{\nu}\SG^{\rho}\SB^{\sigma} = 24\;P_L\;,\;\;
\;\;i\;\epsilon_{\mu\nu\rho\sigma}\;\SB^{\mu}\SG^{\nu}\SB^{\rho}\SG^{\sigma} = -24\;P_R\;.
\EQ
Two other useful identities  are the contractions
\BQ
g_{\mu\nu}\;\SG^{\mu}\SB^{\alpha}\SG^{\nu} = -2 \SG^{\alpha}\;,\;\;g_{\mu\nu}\;\SG^{\mu}(\SB^{\alpha}\SG^{\beta} - \SB^{\beta}\SG^{\alpha})\SB^{\nu} = 0\;,
\\
g_{\mu\nu}\;\SB^{\mu}\SG^{\alpha}\SB^{\nu} = -2 \SB^{\alpha}\;,\;\;g_{\mu\nu}\;\SB^{\mu}(\SG^{\alpha}\SB^{\beta} - \SG^{\beta}\SB^{\alpha})\SG^{\nu} = 0\;.
\EQ
As a final example, the trace of 6 Pauli matrices contains 15 triple $g$ contractions, plus 15 terms in $i\,g\epsilon$, alternating the signs and
conjugating depending on the nature of the epsilon contracted Pauli matrices ($\epsilon\,\SG\SB\SG\SB$ or $\epsilon\,\SB\SG\SB\SG$). The identity looks like 
\BQ
  Tr (\SG_{\lambda}\SB_{\mu}\SG_{\nu}\SB_{\rho}\SG_{\sigma}\SB_{\tau}) = 2 (g_{\lambda\mu}g_{\nu\rho}g_{\sigma\tau}\;-\;g_{\lambda\nu}g_{\mu\rho}g_{\sigma\tau} 
  +  i\;g_{\lambda\mu}\epsilon_{\nu\rho\sigma\tau} - i\;g_{\lambda\nu}\epsilon_{\mu\rho\sigma\tau} +\cdots )
\EQ
The trace of 8 Pauli matrices has 105 term in $g^4$, 210 terms in $gg\epsilon$, and 70 terms in $\epsilon\epsilon$. We recommend using a computer
!
\section{Hodge duality}\label{A3}

In Minkowski space, with signature $(-+++)$, we can define the Hodge dual of a $p$-form ($p=0,1,2,3,4$) as
\BQ
{}{}^*(1) = + \frac{i}{24}\epsilon_{\mu\nu\rho\sigma}\;dx^{\mu}dx^{\nu}dx^{\rho}dx^{\sigma}\;,\\
{}^*(dx_{\mu}) = + \frac{i}{6}\epsilon_{\mu\nu\rho\sigma}\;dx^{\nu}dx^{\rho}dx^{\sigma}\;,\\
{}^*(dx_{\mu}dx_{\nu}) = - \frac{i}{2}\epsilon_{\mu\nu\rho\sigma}\;dx^{\rho}dx^{\sigma}\;,\\
{}^*(dx_{\mu}dx_{\nu}dx_{\rho}) = - i\;\epsilon_{\mu\nu\rho\sigma}\;dx^{\sigma}\;,\\
{}^*(dx_{\mu}dx_{\nu}dx_{\rho}dx_{\sigma}) = +i\;\epsilon_{\mu\nu\rho\sigma}\;,
\EQ
where $\epsilon$ is the totally antisymmetric symbol with $\epsilon_{0123} = 1$
and the (exterior) differentials $dx^{\alpha}$ anticommute with each other.
Our sign choices, the $\pm i$ factors, are explained below.
The definition extends by linearity to a general polyform of mixed degree.
By inspection, we can verify that ${^*}{^*} = 1$. Since the dual of a 2-form is a 2-form,
we can define the self-dual, and the anti-self-dual, projectors
\BE
(P^{\pm})_{\mu\nu\rho\sigma} = \frac{1}{4} (g_{\mu\rho}g_{\nu\sigma}\;-\;g_{\mu\sigma}g_{\nu\rho} \;\mp\;i\;\epsilon_{\mu\nu\rho\sigma})
\;,
\EE
and verify by inspection the that they split the space of 2-forms
\BE
(P^+)^2 = P^+\;,\;\;(P^-)^2 = P^-\;,\;\;P^+P^- = P^-P^+ = 0\;,\;\;P^+ + P^- = 1
\;.
\EE
Let us now consider a superconnection polyform $\AT$ mixing exterior-forms of all degrees:
\BQ
\AT = \phi + a_{\mu}dx^{\mu} + \frac{1}{2}b_{\mu\nu}dx^{\mu}dx^{\nu}
+ \frac{1}{6}c_{\mu\nu\rho}dx^{\mu}dx^{\nu}dx^{\rho}
+\frac{1}{24}e\;\epsilon_{\mu\nu\rho\sigma}dx^{\mu}dx^{\nu}dx^{\rho}dx^{\sigma}\;,
\EQ
where we wrote the 4-form as $e_{\mu\nu\rho\sigma} = e\;\epsilon_{\mu\nu\rho\sigma}$.
Using (C.1), its dual reads 
\BQ
^*\AT = -ie  - \frac{i}{6} c^{\mu\nu\rho} \epsilon_{\mu\nu\rho\sigma}\,dx^{\sigma}  -\frac{i}{4}b^{\mu\nu}\epsilon_{\mu\nu\rho\sigma}\,dx^{\rho}dx^{\sigma} 
\\
+ \frac{i}{6}a^{\mu}\epsilon_{\mu\nu\rho\sigma}\,dx^{\nu}dx^{\rho}dx^{\sigma}
+\frac{i}{24}\phi\epsilon_{\mu\nu\rho\sigma}\,dx^{\mu}dx^{\nu}dx^{\rho}dx^{\sigma}\;.
\EQ
In components, the self-duality condition $\AT = ^*\AT$ expands as
\BQ
\phi = -i e
\;,\;\;a_{\mu} = (^*c)_{\mu} = - \frac{i}{6} \epsilon_{\mu\nu\rho\sigma}\,c^{\nu\rho\sigma}
\;,\\
b_{\mu\nu} = (^*b)_{\mu\nu} = - \frac{i}{2}\epsilon_{\mu\nu\rho\sigma}\,b^{\rho\sigma}\;,
\\
e  = i \phi
\;,\;\;c_{\mu\nu\rho} = (^*a)_{\mu\nu\rho} = - i \epsilon_{\mu\nu\rho\sigma}\,a^{\sigma}\;.
\EQ
By inspection, these equations are consistent, and allow the elimination of $c$ and $e$. Consider now the part of the Dirac operator (4.1) associated to the superconnection,
\BQ
\ATS = \phi + a_{\mu}(\SG^{\mu}+\SB^{\mu})
+ \frac{1}{2}b_{\mu\nu} (\SB^{\mu}\SG^{\nu} + \SG^{\mu}\SB^{\nu})
+ \frac{1}{6}c_{\mu\nu\rho} (\SG^{\mu}\SB^{\nu}\SG^{\rho} + \SB^{\mu}\SG^{\nu}\SB^{\rho})
\\
+\frac{1}{24}e\,\epsilon_{\mu\nu\rho\sigma} (\SB^{\mu}\SG^{\nu}\SB^{\rho}\SG^{\sigma} + \SG^{\mu}\SB^{\nu}\SG^{\rho}\SB^{\sigma}) \;.
\EQ
Using the Pauli matrices identities (A.7), this equation can be rewritten as
\BQ
\ATS = (\phi + i\,e) P_R + (\phi - i\,e) P_L
\\
+ (a_{\mu} - \frac{i}{6}c^{\alpha\beta\gamma}\epsilon_{\alpha\beta\gamma\mu})\,\SG^{\mu}
+ (a_{\mu} + \frac{i}{6} c^{\alpha\beta\gamma}\epsilon_{\alpha\beta\gamma\mu})\,\SB^{\mu}
\\
+ \frac{1}{4}(b_{\mu\nu} + \frac{i}{2} b^{\alpha\beta}\epsilon_{\alpha\beta\mu\nu})\,\SB^{\mu}\SG^{\nu}
+ \frac{1}{4}(b_{\mu\nu} - \frac{i}{2} b^{\alpha\beta}\epsilon_{\alpha\beta\mu\nu})\,\SG^{\mu}\SB^{\nu}
\;.
\EQ
Remarkably, we can recognize in the terms acting on the right Fermions the self-duality conditions (C.6).
Therefore, if $\AT$ is self-dual, then $\ATS$ annihilates the right Fermions:
\BQ
\ATS = ^*\ATS \Rightarrow \ATS \;\psi_R = 0\;,
\\
\ATS \;\psi_L =  (2 \phi + 2 a_{\mu} \SB^{\mu} + \frac{1}{4}(b_{\mu\nu} + *b_{\mu\nu}) \SG^{\mu}\SB^{\nu} )\;\psi_L
\;.
\EQ
\textit{Mutatis mutandis}, if $\AT$ is anti-self-dual, $\ATS$ annihilates the left Fermions.
\BQ
\ATS = - ^*\ATS \Rightarrow \ATS \;\psi_L = 0\;,
\\
\ATS \;\psi_R=  (2 \phi + 2 a_{\mu} \SG^{\mu} + \frac{1}{4}(b_{\mu\nu} - *b_{\mu\nu}) \SB^{\mu}\SG^{\nu} )\;\psi_R
\;.
\EQ

Let us now consider the freedom of the sign choices in the definition of the Hodge dual (C.1).
There are 5 equations.
Asserting that the Hodge duality is an involution (${}^{**} = 1$),
the last 2 signs, for the 3 and 4-forms,
are related to the first 2 choices for the 0 and 1-forms
and the $i$ factors are needed 
in Minkowski space-time (they would disappear in Euclidean space).
We remain with 3 optional choices. The choice for the 2-form
selects which of $b \pm ^*b$ acts on the left spinors, i.e.
is proportional to $\SG\SB$. In quantum field theory this translates
into the orientation of the $\BB B$ propagator: i.e. choosing whether $B$ is emitted by
a left or by a right spinor (section 3). \textit{In fine}, it is a choice of names.
Similarly, the choice of the signs for the 0 and 4-forms
relates to the orientation of the $\PhiB \Phi$ propagator (section 3 and 4).
Finally, the choice of sign for the 1 and 3-forms, joined to
the superchirality constraint $\ATS = ^*\ATS \chi$, correlates
the superalgebra to the spinor helicities. It is a phenomenological choice.
It tells us, in the $SU(m/n)$ case, whether the left or the right spinors
interact with the $SU(m)$ vectors, or reciprocally, whether the left spinors
interact with the $SU(m)$ or the $SU(n)$ vectors. We chose our signs so that
in the lepton/quark model \cite{TM20b}, the left electrons and quarks are $SU(2)$ doublets
and emit $\Phi$ and $B$ fields.

\section{The 't Hooft integrals}\label{A4}
To compute the pole part of the 1-loop divergent Feynman integrals, we need a regularization scheme.
Following, but not exactly, 't Hooft and Veltman diagrammar \cite{diagrammar}, we want to define
dimensional regularization in an axiomatic way just from dimensional analysis and
the linearity of the integrand, and of the integration variable. If we denote $I(f(k))$ a loop
integral of a function $f$ of the momentum $k$, we postulate that

1: UV convergent integrals vanish.

2: The integral $I$ is a linear operator.
\BE
I (af + bg) = a\,I(f) + b\,I(g)\;,\;\;\; a,b \;in\;\CC\;.
\EE

3: We can freely perform a linear change of integration variable
\BE
\int d^dk\;f(k) = \int d^d(ak + b) f(ak + b)\;.
\EE

4: The integral is $d$ dimensional
\BE
\int d^d (ak) f(k) = a^d \;\int d^dk f(k)\;.
\EE
It follows that the integral of a pure power rescales as
\BE
I(k^p) = \int d^dk\,k^p = \int d^d(ak)\,(ak)^p = a^{d+p}  \int d^dk\,k^p = a^{d+p} I(k^p)\;.
\EE
Hence, if $d + p \neq 0$, the integral of a power of $k$ vanishes.
In negative dimension $d = -1$, we recover the Berezin integral
\BE
\int d^{-1}\theta \;\theta = I_1\;.
\EE
In 4 dimensions we obtain the 't Hooft integral
\BE
\int d^4k \frac {1} {(k^2)^2} = I_1\;.
\EE
In the renormalization procedure, we treat $I_1$ as a non standard number. In any equation,
all terms linear in $I_1$ must be isolated and treated separately from the other terms
just like we treat the imaginary part of a complex equation, or the term proportional to
$\sqrt{5}$ in $\sqrt{5}\QQ$ arithmetic.
If in the Lagrangian, the global sum of all the counterterms proportional to $I_1$ cancels out, the theory is called
renormalizable.

Let us now consider the wave function renormalization of a Fermion. Factorizing out
the Dirac matrix and the sign conventions which play no role in the present discussion
we get an integral of the form
\BE
Z(p_{\mu}) = \int d^4k \frac {k_{\mu}} {k^2(k+p)^2}\;.
\EE
To evaluate this integral, we complete the square in the numerator
\BQ
0  = \int d^4k \frac {1} {k^2} = \int d^4k \frac {(k+p)^2} {k^2(k+p)^2} =
\\
= \int d^4k \frac {k^2} {k^2(k+p)^2} + 2p_{\mu}\; \int d^4k \frac {k_{\mu}} {k^2(k+p)^2} + p^2\; \int d^4k \frac {1} {k^2(k+p)^2} =
\\
= 0 + 2\; p_{\mu}\;Z(p_{\mu}) + p^2\;I_1
\EQ
Hence, we conclude that
\BE
Z(p_{\mu}) = \int d^4k \frac {k_{\mu}} {k^2(k+p)^2} = - \frac {1}{2}\; p_{\mu} \;I_1\;.
\EE
More generally, by dimensional analysis, we have 3 kinds of divergent diagrams, a) the Boson Fermion vertices and the 4 Bosons
interactions which are independent of the external momenta; b) the triple Boson vertices and the Fermion propagators which are linear in
the incoming momenta $(p,q)$; c) the Boson propagators which are quadratic in $p$.

Let us first consider the log divergent integral, case a). By dimensional
analysis, if the overall power of $k$ under the integral iz zero, they cannot depend on $p$ or $q$ and must be
of the form
\BE
\int d^4k\; \frac {k_ak_bk_c...k_n}{k^{n+4}} = \alpha (g_{ab} g_{c..} ... + g_{ac} g_{b..} ...)
\EE
where the numerator is symmetric in $(abc...n)$, hence contains all possible pair contractions.
The overall coefficient $\alpha$ is found by tracing all the free indices.
 One gets
\BQ
\int d^4k\; \frac {k_ak_b}{k^6} = \frac {1}{4} g_{ab}\;,\;\;
\\
\int d^4k\; \frac {k_ak_bk_ck_d}{k^8} = \frac {1}{24} (g_{ab}g_{cd} + g_{ac}g_{bd} + g_{ad}g_{bc})\;,\;\;
\\
\int d^4k\; \frac {k_ak_bk_ck_dk_ek_f}{k^{10}} = \frac {1}{192} (g_{ab}g_{cd}g_{ef} + g_{ac}g_{bd}g_{ef} +...\;\; (15\;terms )
\\
\int d^4k\; \frac {k_ak_bk_ck_dk_ek_fk_gk_h}{k^{12}} = \frac {1}{1920} (g_{ab}g_{cd}g_{ef}g_{gh} + ... \;\;(105\;terms )
\EQ

Let us now consider the linear divergent integrals, case b).
By dimensional analysis, the terms of the form $\int d^4k\,k^{2n-3}/k^{2a}(k+p)^{2b}(k+p+q)^{2c}\;,a+b+c = n$, must be of the form $(\alpha p + \beta q)\;I_1$.
To find the coefficient $(\alpha,\beta)$, we differentiate the integrand with respect to $p$ and $q$ and get
\BQ
\int d^4k\; \frac {k_a}{k^2(k+p)^2} = -\frac {1}{2} p_a\;,\;\;
\\
\int d^4k\; \frac {k_ak_bk_c}{k^2(k+p)^2(k+p+q)^2} =  -\frac {1}{12} ((2p+q)_a g_{bc} +(2p+q)_b g_{ca} + (2p+q)_c g_{ab})\;,
\\
\int d^4k\; \frac {k_ak_bk_ck_dk_e}{k^2(k+p)^2(k+p+q)^2(k+p+q+r)^2} =  -\frac {1}{96} ((3p+2q+r)_a g_{bc}g_{de} + ...)\;\;(15 terms)\;,
\EQ
and so on. 
To compute the quadratic Boson propagators, we compute as in a Taylor series half the double derivative of the integrand and obtain
\BQ
\int d^4k\; \frac {k_ak_b}{k^2(k+p)^2} =  \frac {1}{3} (p_ap_b - \frac{1}{4}g_{ab}p^2)\;,\;\;
\\
\EQ
and so on. We verified all the coefficients with a dedicated C-program available on request.

\section {Detailed calculation of the tensor propagator}\label{A5}

In this appendix, we give in detail the quantum field theory evaluation of
the pole-part of the $B$ tensor 2-point function upon inclusion
of a Fermion loop. The hope is to help the interested reader wishing
to understand our notations and reproduce the results.
These calculations are very delicate and feedback would be much appreciated.

We want to compute the pole-part of the diagram

$\;\;\;\;\;\;\;\;\;\;\;\;\;\;\;\;\;\;\;\;\;\;\;\;\;$
\begin{tikzpicture}
\begin{feynman}
\vertex (a) {\(B^i_{\mu\nu}\)};
\vertex [right=of a] (b);
\vertex [right=of b] (c);
\vertex [right=of c] (d){\(\BB^j_{\rho\sigma}\)};
\diagram* {
  (a) -- [gluon] (b),
  (b) -- [anti fermion, half left, edge label =\(\psi_R\) ](c),
  (c) -- [anti fermion, half left, edge label =\(\psi_L\) ] (b),
  (d) -- [gluon] (c),
};
\end{feynman}
\end{tikzpicture}

assuming the Feynman rules

\begin{tikzpicture}
\begin{feynman}
  \vertex (a1) {\(\psi_L\;\;=\;p^{\mu}\SB_{\mu}/p^2\)};
  \vertex [left = of a1] (a);
\vertex [left = of a] (x){\(\overline{\psi_L}\)};
\diagram* {
  (x) --  [fermion](a),
};
\end{feynman}
\end{tikzpicture}
$\;\;\;\;\;\;\;\;\;\;\;$
$\;\;\;\;\;\;\;\;\;\;\;$
\begin{tikzpicture}
\begin{feynman}
  \vertex (a1) {\(\psi_R\;\;=\;p^{\mu}\SG_{\mu}/p^2\)};
  \vertex [left = of a1] (a);
\vertex [left = of a] (x){\(\overline{\psi_R}\)};
\diagram* {
  (x) --  [fermion](a),
};
\end{feynman}
\end{tikzpicture}

\begin{tikzpicture}
\begin{feynman}
  \vertex (a1) {\(\BB^i_{\mu\nu}=\lX_i\;\frac{1-\chi}{4}\;\SG_{\mu}\SB_{\nu}\)};
  \vertex [left = of a1] (a);
\vertex [left = of a] (x);
\vertex [below left=of x] (b){\(\psi_L\)};
\vertex [above left=of x] (c){\(\overline{\psi_R}\)};
\diagram* {
  (a) -- [gluon] (x),
  (x) --  [anti fermion](b),
  (x) --  [fermion](c),
};
\end{feynman}
\end{tikzpicture}
$\;\;\;\;\;\;\;\;\;\;\;$
\begin{tikzpicture}
\begin{feynman}
  \vertex (a1) {\(B^i_{\mu\nu}=\lX_i\;\frac{1+\chi}{4}\;\SB_{\mu}\SG_{\nu}\)};
  \vertex [left = of a1] (a);
\vertex [left = of a] (x);
\vertex [below left=of x] (b){\(\psi_R\)};
\vertex [above left=of x] (c){\(\overline{\psi_L}\)};
\diagram* {
  (x) -- [gluon] (a),
  (x) --  [anti fermion](b),
  (x) --  [fermion](c),
};
\end{feynman}
\end{tikzpicture}

Notice that we do not need to know in advance the Feynman rule for the
propagator of the $B$ field to compute its counterterm. We find
\BQ
Z^{ij}_{\mu\nu\rho\sigma} = \frac{1}{4}\;\int {d^4k \frac {k^{\alpha}(k+p)^{\beta}}{k^2(k+p)^2}}\;
Tr (\SG_{\alpha}\SB_{\mu}\SG{_\nu}\SB_{\beta}\SG_{\rho}\SB_{\sigma})
\;Tr(\lX^i\;\frac{1 - \chi}{2}\;\lX^j\;\frac{1 + \chi}{2})\;.
\EQ
Since $\chi$ satisfies $\chi^2 = 1$ and anticommutes with the odd matrices $\lX_i$ and $\lX_j$, the charge trace can be rewritten as (3.7):
\BE
\kappa_{ij} = Tr(\lX_i\;\frac{1 - \chi}{2}\;\lX_j\;\frac{1 + \chi}{2}) = \frac{1}{2} Tr(\lX_i\lX_j) + \frac{1}{2}STr(\lX_i\lX_j)\;.
\EE
As discussed in section 3, the $\kappa_{ij}$ propagator hesitates between a Lie algebra $Trace$ and a Kac superalgebra $Supertrace$.
The ambiguity is only lifted in section 6 when the precise Fermion content of the model is taken into account.

Using (D.11) and (D.13), the pole-part of the integral over the momenta gives:
\BE
\int {d^4k \frac {k_{\alpha}(k+p)_{\beta}}{k^2(k+p)^2}}\;
=  \frac {1}{3} (p_{\alpha}p_{\beta} - \frac{1}{4}g_{\alpha\beta}p^2) - \frac{1}{2} p_{\alpha}p_{\beta}\;
=  -\frac {1}{6} p_{\alpha}p_{\beta} - \frac{1}{12}g_{\alpha\beta}p^2\;.
\EE
Finally the trace over the six Pauli matrices is the most complicated term (B.9), but it can be simplified.
Observe first that the tern in $g_{\alpha\beta}$ in (E.3) can be dropped thanks to
the last identity (B.8) and the fact that $B_{\mu\nu}$ is an antisymmetric tensor. In the same way
the terms proportional to $g_{\mu\nu}$ and to $g_{\rho\sigma}$ in the trace of the six Pauli
matrices, a generalization of (B.6),  can be dropped and the terms proportional to the epsilon symbol can be dropped
if they contain $\alpha\beta$ or can be absorbed by the $\BB B$ fields if they
contain $\mu\nu$ or $\rho\sigma$. We are left with 6 equivalent contractions,
4 coming from the $ggg$ trace, and 2 coming from $g\epsilon$, compensating the
factor $1/6$ coming from the loop integral (E.3). Finally
\BE
Z^{ij}_{\mu\nu\rho\sigma} = - g^{ij}\;P^-_{\mu\nu\alpha\beta}\;p^{\alpha}p^{\gamma}g^{\beta\delta}P^+_{\gamma\delta\rho\sigma}
\;,\EE
our equation (4.7), which is equivalent to (3.6), as the $P^{\pm}$ projectors (C.2)
can be absorbed by the $(\BB, B)$ fields.

The calculation of the Feynman diagrams leading to (4.6) and (4.8)
are analogous, with traces involving up to 8 Pauli matrices and a flurry of $\epsilon$ symbols which
can be all reabsorbed in the self-duality of the $\BB B$ fields. The loop integral generates several terms
like in (E.3), summing up to the only Lorentz invariant contraction $F^{\mu\nu}\BB_{\mu\nu}$
which is equivalent to $i\,\epsilon^{\mu\nu\rho\sigma}F_{\mu\nu}\BB_{\rho\sigma}$ since $\BB$ is self-dual.


\end{document}